# Warkworth 12-m VLBI Station: WARK12M

*Sergei Gulyaev, Tim Natusch, Stuart Weston, Neville Palmer, David Collett*

**Abstract**

This report summarizes the geodetic VLBI activities in New Zealand in 2010. It provides geographical and technical details of WARK12M - the new IVS network station operated by the Institute for Radio Astronomy and Space Research (IRASR) of Auckland University of Technology (AUT). The details of the VLBI system installed in the station are outlined along with those of the collocated GNSS station. We report on the status of broadband connectivity and on the results of testing data transfer protocols; we investigate UDP protocols such as 'tsunami' and UDT and demonstrate that the UDT protocol is more efficient than 'tsunami' and 'ftp'. In general, the WARK12M IVS network station is fully equipped, connected and tested to start participating in regular IVS observational sessions from the beginning of 2011.

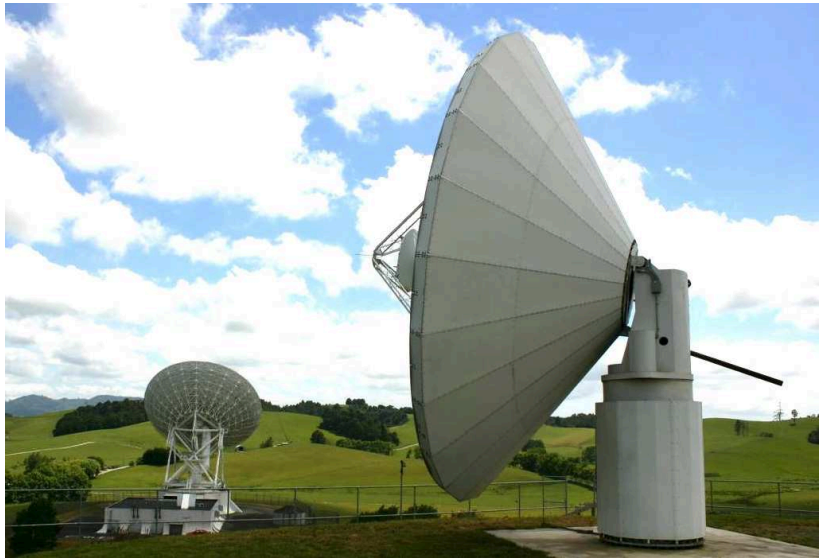

Figure 1. Warkworth 12-m (right) and Warkworth 30-m (left) antennas. Photo: Jordan Alexander

## 1. Introduction

The New Zealand 12-m radio telescope is located some 60 km north of the city of Auckland, near the township of Warkworth (see Figure 1). It was manufactured by Patriot Antenna Systems (now Cobham Antenna Systems), USA. The antenna specifications are provided in Table 1. The radio telescope is designed to operate at S and X bands and supplied with an S/X dual-band dual-polarisation feed. The antenna is equipped with a digital base band converter (DBBC) developed by the Italian Institute of Radio Astronomy, Symmetricom Active Hydrogen Maser MHM-2010 (75001-114) and with a Mark5B+ data recorder developed at MIT Haystack Observatory.

The support foundation for the antenna is a reinforced concrete pad that is 1.22 m thick by $6.7 \times 6.7$ meters square. The ground that the foundation is laid on consists of weathered sandstone/mudstone, i.e. of sedimentary origin, laid down in the Miocene period some 20 million years ago. The pedestal is essentially a steel cylinder of 2.5 m diameter. It supports the antenna elevation axis which is at a height of approximately 7.1 m above ground level. Apart from the pedestal all other components of the antenna (the reflector and feed support structure) are constructed of aluminium. The radio telescope is directly connected to the regional network KAREN (Kiwi Advanced Research and Education Network), which provides connectivity between New Zealand's educational and research institutions [1].

Table 1. Specifications of the Warkworth 12-m antenna

| | |
|---|---|
| Antenna type | Fully-steerable, dual-shaped Cassegrain |
| Manufacturer | Cobham/Patriot, USA |
| Main dish Diam. | 12.1 m |
| Secondary refl. Diam. | 1.8 m |
| Focal length | 4.538 m |
| Surface accuracy | 0.35 mm |
| Pointing accuracy | 18″ |
| Frequency range | 1.4—43 GHz |
| Mount | alt-azimuth |
| Azimuth axis range | 90° ± 270° |
| Elevation axis range | 4.5° to 88° |
| Azimuth axis max speed | 5°/s |
| Elevation axis max speed | 1°/s |
| Main dish F/D ratio: | 0.375 |

## 2. Antenna position survey

The reference point of a VLBI site is defined as the intersection of the azimuth and elevation axes of the telescope. A preliminary survey has been conducted in collaboration with the New Zealand Crown Research Institute (CRI), GNS Science and Land Information New Zealand (LINZ) to determine an initial estimate of the reference point of the VLBI site WARK12M (see details in [2]). A real-time kinematic (RTK) GPS method was used to derive the position with respect to the collocated GPS station WARK.

The GPS station WARK was established in November 2008 at the radio telescope site and is one of thirty nine PositioNZ network stations in New Zealand [3]. All data received from the PositioNZ stations are compiled into 24 hour sessions and are processed to produce daily positions for each station in terms of ITRF2000. The coordinates for WARK used in the following calculations were derived by averaging the daily coordinate solutions for February 19 through March 09, 2010. The RTK reference receiver was set up in an arbitrary location with clear sky view and was configured to record raw observations in addition to transmitting real-time corrections. The RTK station was later post-processed with respect to WARK and all RTK rover surveyed positions were subsequently adjusted relative to the updated reference position.

Several points on the rim of the main reflector were identified and each point was measured several times with the RTK rover while the telescope was repositioned in elevation and azimuth between each measurement. The rover GPS antenna was mounted on a 0.5 m survey pole and was hand held for each measurement. Access to the rim of the reflector was achieved with a hydraulic cherry picker.

The following sequence of observation for determination of the horizontal axis was applied. The telescope azimuth axis was held fixed (nominally 0°). A point near the highest edge of the reflector was identified and measured with the telescope in four positions of elevation, from almost zenith (88°) to as high as the cherry picker could reach (38°). This was repeated with a point identified on the edge of the reflector to one side of the telescope. Five positions of elevation were measured at this point, from 10° to 80°.

The observation sequence for determination of the vertical axis was as follows. The telescope elevation axis was held fixed (nominally 80°). Three points around the edge of the reflector were identified. The telescope azimuth axis was rotated into three positions such that each identified point could be measured consecutively from the surveyor's location in the cherry picker. The cherry picker was then repositioned twice around the perimeter of the telescope and the measurements were repeated at each cherry picker location. This provided three measurements of each identified point with a fixed telescope zenith and varying azimuths. The resulting points from these measurements describe two vertical circles of rotation which define the movable elevation axis and three horizontal circles of rotation which define the fixed azimuthal axis. The coordinates for all subsequent calculations were retained as geocentric Cartesian coordinates to avoid any possibility of errors related to transformation of projection.

To determine the axes and their intersection point, the equation of a circle from three points was used to calculate all possible combinations of three observed points which define a circle of rotation. Mean values were taken for all horizontal axis definitions and all vertical axis definitions. The midpoint of the closest point of approach of each axis with the other was used as the final estimate of the point of intersection. The distance between the axes was calculated to be 24 mm. Based on the variation of results for different combinations of survey points, we estimate that the accuracy of the determined intersection point is within 0.1 m.

In summary, the following coordinates of the intersection of the azimuth and elevation axes for the radio telescope WARK12M were derived in terms of ITRF2000 at the epoch of the survey (March 2010):

$$\begin{aligned} X &= -5115324.5 \pm 0.1 \text{ m} \\ Y &= 477843.3 \pm 0.1 \text{ m} \\ Z &= -3767193.0 \pm 0.1 \text{ m} \end{aligned} \qquad (1)$$

It is intended that the radio telescope reference point coordinates will subsequently be re-determined to a higher accuracy with the use of a variety of terrestrial and GNSS survey techniques (e.g. [4]) and a more rigorous least squares analysis of the observations. Four geodetic survey monuments have been built within 15–20 m from the antenna pedestal for this purpose.

## 3. Network connectivity and data transfer protocols

Internationally, New Zealand's major broadband supplier is Southern Cross Cables Ltd, a commercial organisation, which owns and operates multi-wavelength fibre cables with total capacity of about 2 Tbps connecting New Zealand with Australia in the west and with the USA in the northeast directions. Locally, the regional advanced network operating in New Zealand is KAREN network, which provides 10 Gbps connectivity between New Zealand's educational and research institutions. In April 2010, KAREN established a GigaPoP at the Warkworth Radio Astronomical Observatory, which provides a 1 Gbps connectivity for the WARK12M radio telescope. International connectivity was initially limited to 155 Mbps (New Zealand – Australia) and 622 Mbps (New Zealand – USA). By the end of 2010 both international directions were upgraded to 1 Gbps.

With the connection of WARK12M to the KAREN network our intention is to optimize the transfer of large volumes of IVS observational data to correlation centres in Europe (Bonn), North America (Washington) and Australia (Perth). In this section we discuss the use of FTP over TCP/IP protocol for transferring data, and compare the performance of 'tsunami' and UDT (UDP-based Data Transfer) via the network protocol UDP.

Table 2 presents the destinations (column 1) that connectivity was achieved with in 2010, the protocols that were verified for data transfer (columns 2 and 3) and command line access to remote servers for initiating data transfers (column 4).

Table 2. Connectivity established via KAREN in 2010

| Destination | Protocol | | Command | Date |
|---|---|---|---|---|
| CSIRO (Australia) | UDP | Tsunami, UDT | ssh | 01/04/2010 |
| Bonn (Germany) | UDP | Tsunami, UDT | ssh | 01/06/2010 |
| JIVE (Netherlands) | UDP | – | iperf | 27/07/2010 |
| Metsahovi (Finland) | UDP | Tsunami, UDT | ssh | 21/07/2010 |

'Tsunami' is an UDP file transfer protocol developed by Jan Wagner of the Metsahovi Radio Observatory in 2007 [5]. This is the protocol of choice for sending files to Bonn for IVS observations, as stipulated by the MPIfR. Another UDP protocol called UDT was developed at the University of Illinois in 2005 [6, 7]. UDT has some advantages over 'tsunami'. For example, UDT has an application programming interface (API) allowing easy integration with existing or future applications. Also, UDT is a better citizen on the network leaving bandwidth for TCP and other UDP protocols, a capability which is very important on a shared network such as KAREN.

The results of data transfer tests between the IRASR and data processing centre in Bonn are presented in Table 3. They were obtained by transferring an actual 16 bit VLBI file produced in observations with the 12-m radio telescope. The data was transferred from the IRASR's IBM Blade server via the KAREN network using the default settings for each protocol with no tuning. Column 1 shows the protocol used, column 2 gives the amount of data sent in bytes, column 3 provides the time it took to transfer the data and column 4 shows an average throughput rate (each test was repeated 5–10 times).

Table 3 clearly demonstrate the advantage of the UDT protocol over 'tsunami' and 'ftp'. It is more than two times faster than 'tsunami' and more than four times faster than the standard 'ftp' protocol. Tests were conducted repeatedly over several days and at different times resulting in slightly different average rates without changing the main conclusion that the UDT protocol is superior to 'ftp' and 'tsunami'. A traceroute command from the AUT Blade centre to the IP address in Bonn gave a route of 14 hops. Re-issuing this command repeatedly over several months showed that the route appears stable, without any changes. A high number of hops on the route

Table 3. Data transfer statistics: IRASR to Bonn

| Protocol | Bytes | Time (s) | Throughput (Mbps) |
|----------|-------|----------|-------------------|
| Ftp      | 65 G  | 8016     | 65                |
| Tsunami  | 65 G  | 3466     | 151               |
| UDT      | 65 G  | 1920     | 273               |

(14) demonstrates the complexity of the path and explains why data transfers via protocols such as 'ftp' are not efficient.

In June 2010 file transfer tests to the correlator site at Curtin University, Perth were conducted. Using 'tsunami', a rate of 300 Mbps was achieved, while UDT was superior at 400 Mbps sustained throughput. In December 2010 the first eVLBI tests from the Mk5B recorder at the Observatory to a server at CSIRO in Australia using UDP were conducted. The required data rate of 512 Mbps was achieved sustainably.

It is worth mentioning a few more data transfer sessions conducted in 2010. In August 2010 WARK12M participated in observations of the ESA's Mars Express (MEX) spacecraft orbiting Mars. Data (86 GB in total) was transmitted to Metshovi via KAREN using the UDP protocol 'tsunami' immediately after the observational session. The next set of MEX observational data totalling 187 Gbytes was sent to Metshovi using UDT. Average rates sustained were 250 and 300 Mbps respectively.

Another set of experiments were conducted in September 2010 aiming to test the 'tsunami' protocol for streaming VLBI data directly from the radio telescope receiving system (PCEVN) via KAREN to Metshovi. This test was an important step towards real-time eVLBI.

## 4. Warkworth 30-m antenna: WARK30M

In November 2010 Telecom New Zealand handed over to Auckland University of Technology its 30-metre Cassegrain wheel-and-track beam-waveguide antenna. It was manufactured in 1984 by Nippon Electric Corp., and since then it was used by Telecom NZ for communication between New Zealand and various Pacific Islands. The antenna is located near Warkworth, just 200 metres north of WARK12M antenna (see Figure 1). According to expert opinion, the antenna is in good state, and after conversion to a radio telescope it has the potential to contribute to both astronomical and geodetic VLBI.